\documentclass
  {article}
\usepackage{graphicx}
\newcommand{\lora} {\boldmath$\longrightarrow$}

\title{Second Law in Classical Non-Extensive
  Systems\footnote{presented at the First International Conference on
    Quantum Limits to the Second Law, San Diego 2002 }}

\author{D.H.E. Gross\\
Hahn-Meitner Institute and Freie Universit{\"a}t Berlin,\\
Fachbereich Physik.\\ Glienickerstr. 100\\ 14109 Berlin, Germany}

\begin{document}

\maketitle

\begin{abstract}
  Equilibrium statistics of Hamiltonian systems is correctly described
  by the microcanonical ensemble, whereas canonical ones fail in the
  most interesting, mostly inhomogeneous, situations like phase
  separations or away from the ``thermodynamic limit'' (e.g.
  self-gravitating systems and small quantum
  systems)~\cite{gross189,gross174,gross186}. A new derivation of the
  Second Law is presented that respects these fundamental
  complications. Our ``geometric foundation of
  Thermo-Statistics''~\cite{gross186} opens the fundamental
  (axiomatic) application of Thermo-Statistics to non-diluted systems
  or to ``non-simple'' systems which are not similar to (homogeneous) fluids.
  Supprisingly, but also understandably, a so far open problem c.f.
  ref.~\cite{uffink01} page 50 and page 72.
\end{abstract}


\section{Introduction}
Phase separation of normal systems and also the equilibrium of closed
non-extensive systems are not described by the canonical and
grand-canonical ensembles. Only the microcanonical ensemble based on
Boltzmann's principle $S=\ln{W}$, with $W(E)=\epsilon_0
tr[\delta(E-H)]$, the classical or quantum number of states, describes
correctly the unbiased uniform filling of the energy-shell in
phase-space.

The various ensembles are equivalent only if the system is infinite
and homogeneous, i.e. in a pure phase. As this conference addresses
the Second Law in quantum systems (i.e. small systems) and moreover,
as most systems in nature are inhomogeneous it seems necessary to
take the above complication serious. We should deduce the Second Law
from reversible mechanics without invoking the thermodynamic limit
($\lim_{V\to\infty, N/V=\rho}$) and using the microcanonical ensemble
with sharply defined conserved quantities.
\section{Boltzmann's principle, the microcanonical ensemble}
The key quantity of statistics and thermodynamics is the {\em entropy}
$S$.  Its most fundamental definition is as the
logarithm of the {\em area} $W(E)$ of the manifold $\cal{E}$ in the
N-body phase-space given by Boltzmann's principle~\cite{einstein05d}:

\begin{eqnarray}
W(E,N,V)&=&\epsilon_0 tr\delta(E-H_N)\nonumber\\
tr\delta(E-H_N)&=&\int{\frac{d^{3N}p\;d^{3N}q}{N!(2\pi\hbar)^{3N}}
\delta(E-H_N)}.\label{wenv}
\end{eqnarray}
\begin{equation}
\fbox{\fbox{S=k~\mbox{ln}W}}\label{boltzmentr1}
\end{equation}
{\em Boltzmann's principle is the only axiom necessary for
  thermo-statistics.}  With it Statistical Mechanics and also
Thermodynamics become {\em geometric} theories\cite{gross186}. For
instance all kinds of phase-transitions are entirely determined by
{\em topological} peculiarities of the ($6N-1$)-dim. manifold
${\cal{E}}(E,N,\cdots)$ of all points in the $6N$-dim.  phase space
with given energy $E$, particle number $N$ etc.  More precisely, the
topology of $S(E,N,\cdots)$ determines the thermodynamic properties of
the system including all its
phase-transitions~\cite{gross189,gross174}.

\section{Approach to equilibrium, Second Law}
\subsection{Zermelo's paradox}
When Zermelo~\cite{zermelo97} argued against Boltzmann, that following
Poncarr\'e any many-body system must return after the Poincarr\'e
recurrence time $t_{rec}$ and consequently its entropy cannot always
grow, but must decay also, Boltzmann~\cite{boltzmann1896} answered
that for any macroscopic system $t_{rec}$ is of several orders of
magnitude larger than the age of the universe, c.f.
Gallavotti~\cite{gallavotti99}.  Still today, this is the answer given
when the Second Law is to be proven microscopically,
c.f.~\cite{gaspard97}. I.e. the thermodynamic limit seems necessary
for the validity of the Second Law~\cite{lebowitz99,lebowitz99a}.
Then, however, Zermelo's paradox becomes blunted.

However, the reason for the Second Law is not the large recurrence
times but the probabilistic nature of Thermodynamics.  Here, I argue,
even a {\em small} system approaches equilibrium with a rise of its
entropy $\Delta S\ge 0$ under quite general conditions.  Thus,
Zermelo's objection must be considered much more seriously.

However, care must be taken, Boltzmann's definition of entropy
eq.(\ref{boltzmentr1}) is only for systems at equilibrium.  To be
precise: in the following I will consider the equilibrium manifold
${\cal{M}}(t)={\cal{E}}(E,V_a)$ at $t\le t_0$.  At $t_0$ the
macroscopic constraint $V_a$ is quickly removed e.g.  a piston pulled
quickly out to $V_a+V_b$, and the ensemble is followed in phase-space
how it approaches the new equilibrium manifold ${\cal{E}}(E,V_a+V_b)$
see fig.(\ref{spaghetti}).
\subsection{The solution}
Due to the reduced, incomplete, description of a $N$-body system by
Thermodynamics does entropy not refer to a single point in $N$-body
phase space but to the whole ensemble ${\cal{E}}$ of points.  It is
the $\ln(W)$ of the geometrical size $W$ of the ensemble. Every
trajectory starting at different points in the initial manifold
${\cal{M}}(t=t_0)={\cal{E}}(E,V_a)\in{\cal{E}}(E,V_a+V_b)$ spreads in
a non-crossing manner over the available phase-space
${\cal{E}}(E,V_a+V_b)$ but returns after $t_{rec}$.  Different points
of the manifold, or trajectories, have different $t_{rec}$ which are
normally incommensurable. I.e. the manifold ${\cal{M}}(t>t_0)$ spreads
irreversibly over ${\cal{E}}(E,V_a+V_b)$. ${\cal{M}}(t)$ will never go
back into the volume $V_a$.
\subsubsection{Mixing}
When the system is dynamically mixing then the manifold ${\cal{M}}(t)$
will ``fill'' the new ensemble ${\cal{E}}(E,V_a+V_b)$. Though at
finite times the manifold remains compact due to Liouville and keeps
the volume $W(E,V_a)$, but as already argued by
Gibbs~\cite{gibbs02,gibbs36} ${\cal{M}}(t)$ will be filamented like
ink in water and will approach any point of ${\cal{E}}(E,V_a+V_b)$
arbitrarily close. Then, $\lim_{t\to\infty}{\cal{M}}(t)$
becomes dense in the new, larger ${\cal{E}}(E,V_a+V_b)$.
\subsubsection{Macroscopic resolution, fractal distributions and
  closure~\cite{gross183}}

Also an idealized mathematical analysis is possible here: Statistical
Mechanics as a probabilistic theory describes all systems in the same
volume $V$ and with the same few conserved control parameters like
energy $E$, particle number $N$ etc. It is unable to distinguish the
points $\in M(t)$ from the neighboring points in its closure
$\overline{{\cal{M}}(t)}$. Entropy is also the $\ln$ of the size of
$\overline{{\cal{M}}(t)}$. The {\em closure}
$\overline{{\cal{M}}(t=\infty)}$ becomes equal to
${\cal{E}}(E,V_a+V_b)$. I.e. the entropy $S(t=\infty)>S(t_0)$. {\em
  This is the Second Law for a finite system.}

\begin{figure}[!]
\hspace*{1.3cm}\\
\begin{minipage}[h]{6cm}
\begin{center}$V_a$\hspace{2cm}$V_b$\end{center}
\vspace*{0.2cm}
\includegraphics[bb=0 0 404 404, angle=-0,
width=5.5cm,
clip=true]{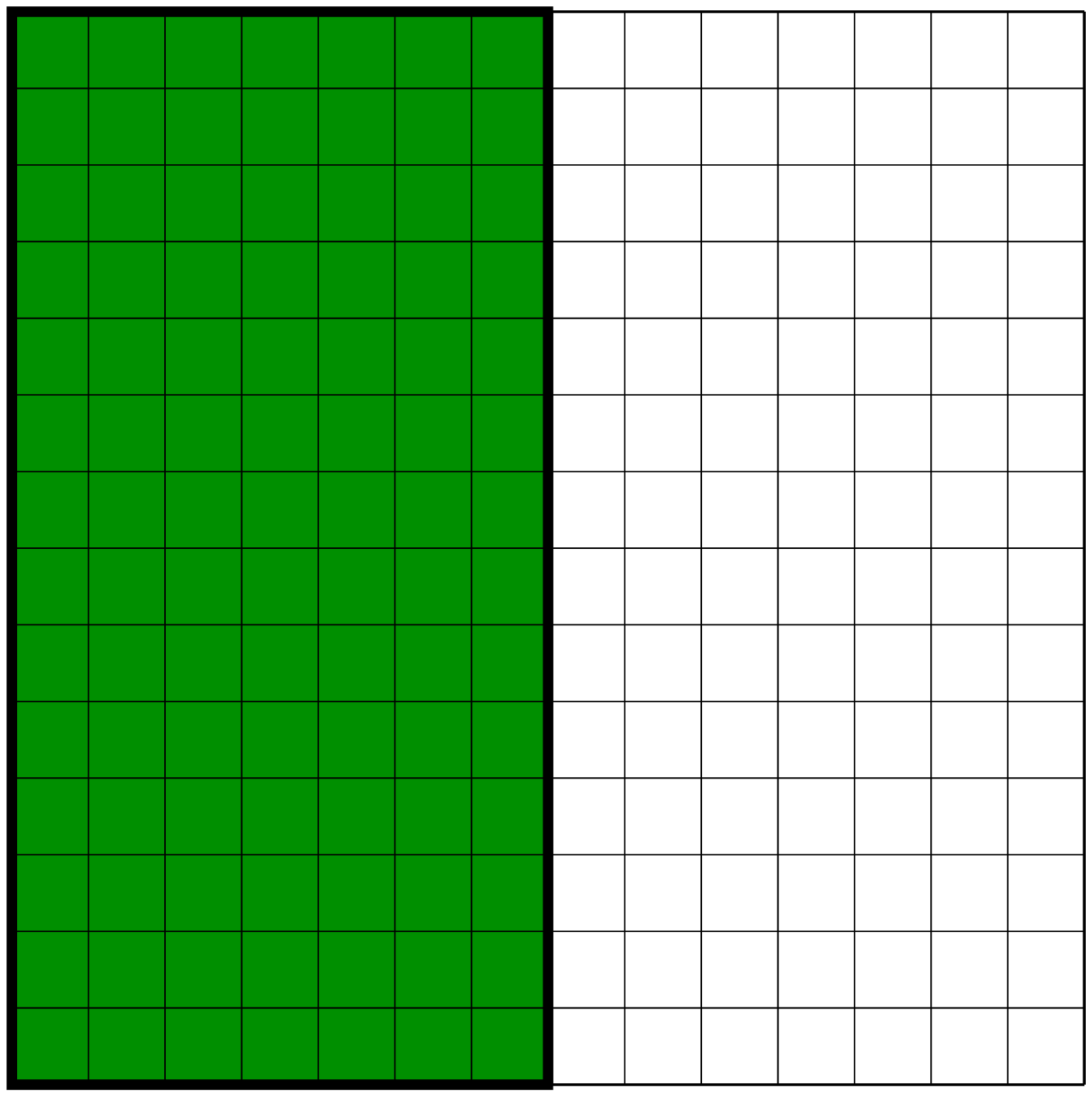}\begin{center}$t<t_0$\end{center}
\end{minipage}\lora\begin{minipage}[h]{6.5cm}
\begin{center}$V_a+V_b$\end{center}\vspace{-0.8cm}
\includegraphics[bb=0 0 490 481, angle=-0, width=6.5cm,
clip=true]{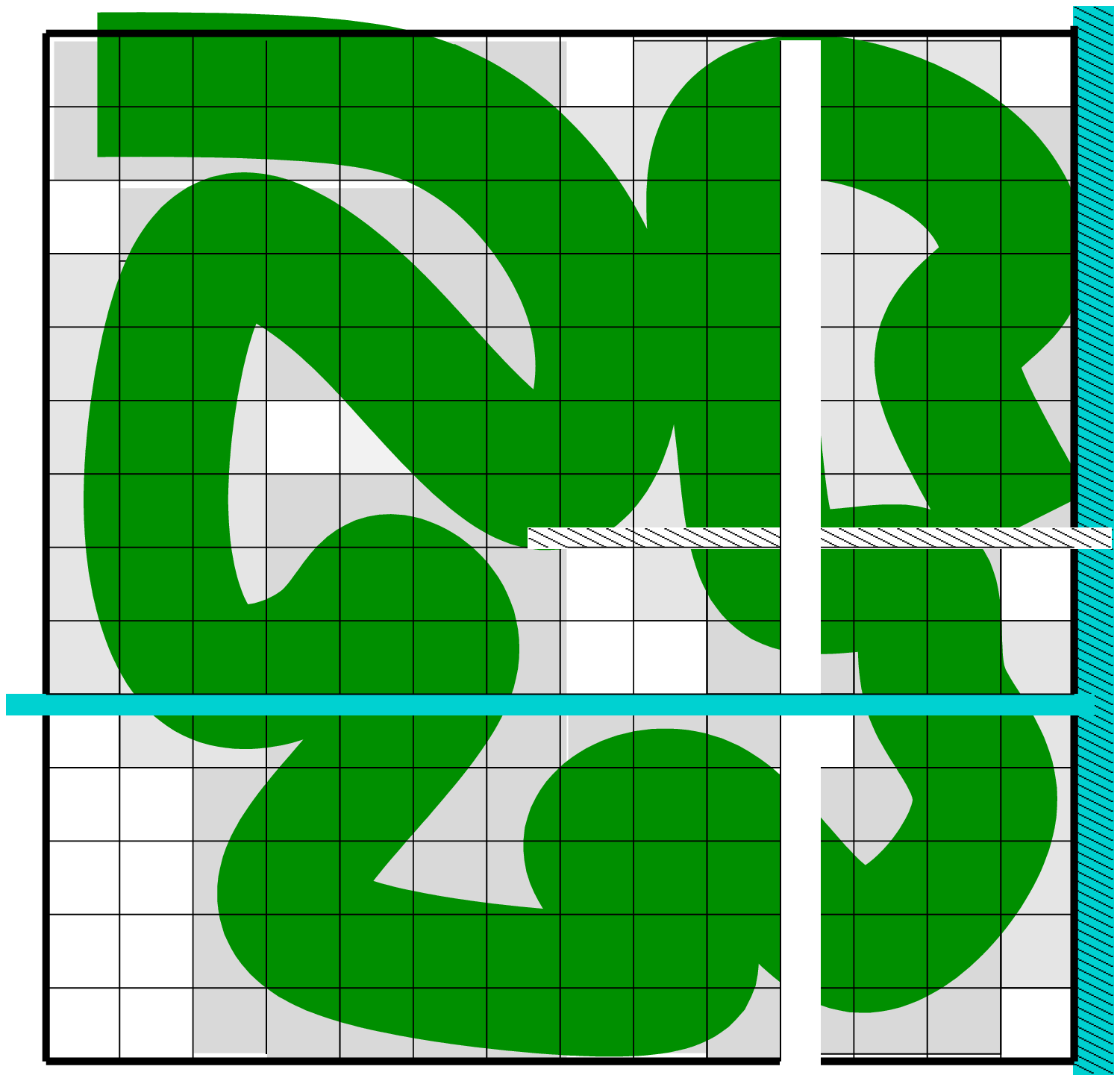}\begin{center}$t>t_0$\end{center}
\end{minipage}\\
\caption{The compact set ${\cal{M}}(t_0)={\cal{E}}(V_a)$, left
side,
  develops into an increasingly folded but non-crossing
  ``spaghetti''-like distribution ${\cal{M}}(t)$ in the phase-space
  with rising time $t$ after opening the volume $V_b$. The right
  figure shows only the early form of the distribution. At much later
  times it will become more and more fractal and finally dense in the
  new larger phase space.  The grid illustrates the boxes of the
  box-counting method.  All boxes which overlap with ${\cal{M}}(t)$
  contribute to the box-counting volume and are shaded gray.  Their
  number is $N_\delta$
  \label{spaghetti}}
\end{figure}

We calculate the closure of the ensemble by box
counting~\cite{falconer90}.  Here the phase-space is divided in
$N_\delta$ equal boxes of volume $\delta^{6N}$. The number of boxes
which overlap with ${\cal{M}}(t)$ is $N_\delta$ and the box-counting
volume is then:
\begin{eqnarray}
\Omega_d(\delta)&=&N_\delta \delta^d\\
&&\mbox{here with $d=6N-1$}\nonumber\\
\Omega_d&=&\lim_{\delta\to 0}\Omega_d(\delta).
\end{eqnarray}
The box-counting method is illustrated in fig.(\ref{spaghetti}).  The
important aspect of the box-counting volume of a manifold is that it
is equal to the volume of its closure.

At finite times ${\cal{M}}(t)$ is compact. Its volume $W(t)$ equals that of
its closure $\equiv W(t_0)$. However,
calculated with finite resolution $\delta>0$, $W_{\delta}(t)$
becomes $\ge W(t)$ for $t$ larger than some $t_\delta$, where
\begin{eqnarray}
W_{\delta}(t)&=&\nonumber\\
\lefteqn{\hspace{-3cm}
\displaystyle{B_d^\delta
\hspace{-0.5 cm}\int}\;\;\frac{d^{3N}p_t\;d^{3N}q_t}{(2\pi\hbar)^{3N}}
\;\delta\left(E-H_N\{q(t),p(t);\}\right)}\\
\lefteqn{\hspace{-3cm}\mbox{with the constraint }
[q(t_0),p(t_0)\in {\cal{E}}(V_a)]}\nonumber\\
\lefteqn{\hspace{-3cm}\mbox{and the definition:}}\nonumber\\
\displaystyle{B_d^\delta\hspace{-0.5 cm}\int}{\;\;f(q)d^dq}&:=&\Omega_d(\delta)
\overline{f(q)}\nonumber.
\end{eqnarray}
A natural finite resolution is given by Quantum Mechanics:
\begin{equation}
\delta=\sqrt{2\pi\hbar},
\end{equation}
but Thermodynamics allows a much coarser resolution because of the
insensitivity of the usual macroscopic observables. Then the
equilibration time $t_\delta$ will also be much shorter.
Thus we should reformulate Boltzmann's principle
eqs.~(\ref{boltzmentr1}) as:
\begin{equation}
S(t)=\ln(W_\delta(t)),\label{boltz_box}
\end{equation}
or mathematically correct, though unphysical, at infinite times:
\begin{eqnarray}
S(t\to\infty)&=&
\lim_{\delta\to 0}\lim_{t\to\infty}{\ln(W_\delta(t))}\nonumber\\
&=&N\ln[(V_a+V_b)/V_a]+S(t_0),
\end{eqnarray}
which agrees with the old one eq.~(\ref{boltzmentr1}) for
compact manifolds like ${\cal{E}}$.
\section{Conclusion}

The {\em geometric interpretation of classical equilibrium Statistical
  Mechanics}~\cite{gross186} by Boltzmann's principle offers an
extension also to the equilibrium of non-extensive systems. In more
fundamental, axiomatic terms, it opens the application of
Thermo-Statistics to ``non-simple'' systems which are not similar to
fluids or systems in contact with an ideal gas.  Surprisingly, but 
also understandably, this is a so far
open problem c.f. ref.~\cite{uffink01} page 50 and page 72.

Because microcanonical Thermodynamics as a macroscopic theory controls
the system by a few, usually conserved, macroscopic parameters like
energy, particle number, etc. it is an intrinsically probabilistic
theory. It describes all systems with the same control-parameters
simultaneously. If we take this seriously and avoid the so called
thermodynamic limit ($\lim_{V\to\infty, N/V=\rho}$), the theory can be
applied to small systems like the usual quantum systems addressed in
this conference but even to the really large, usually {\em
  inhomogeneous}, self-gravitating systems, c.f.\cite{gross187}.

Within the new, extended, formalism several principles of traditional
Statistical Mechanics turn out to be violated and obsolete. E.g. at
phase-separation heat (energy) can flow from cold to
hot~\cite{gross189}.  Or phase-transitions can be classified
unambiguously in astonishingly small systems. These are by no way
exotic and wrong conclusions. On the contrary, many experiments have
shown their validity.  I believe this approach gives a much deeper
insight into the way how many-body systems organize themselves than
any canonical statistics is able to.  The thermodynamic limit clouds
the most interesting region of Thermodynamics, the region of
inhomogeneous phase-separation.

Relevant for most of the contributions to this conference is the fact
that the various canonical ensembles are not equivalent to the
fundamental microcanonical one. They fail to describe the equilibrium
of small systems like quantum systems, as well the largest systems
possible like self-gravitating ones, or the thermodynamical most
important situations like phase-separation with their negative
heat-capacity~\cite{gross189}.

Because of the only {\em one} underlying axiom, Boltzmann's principle
eq.(\ref{boltzmentr1} or \ref{boltz_box}), the {\em geometric
  interpretation}~\cite{gross186} keeps statistics most close to
Mechanics and, therefore, is most transparent. The Second Law ($\Delta
S\ge 0$) is shown to be valid in {\em closed, small} systems under
quite general dynamical conditions.

One may consider this as an artificial mathematical construct, far
away from our daily experience about temperature, heat etc.  However,
it must be emphasized that in the usual macroscopic, extensive world
nothing changes. The change of entropy is still the change of heat
over temperature: $dS=dQ/T$. However, the application of
Thermodynamics to unusual systems like quantum systems or
non-extensive systems demands an abstract mathematical extension like
the one proposed here.




\begin{thebibliography}{10}

\bibitem{gross189}
Gross, D., ``Thermo-Statistics or Topology of the Microcanonical
Entropy
  Surface'', in \emph{Dynamics and Thermodynamics of Systems with Long Range
  Interactions}, edited by T.Dauxois, S.Ruffo, E.Arimondo, and M.Wilkens,
  Springer, Heidelberg, 2002, Lecture Notes in Physics, pp.
  21--45,cond--mat/0206341.

\bibitem{gross174}
Gross, D., \emph{Microcanonical thermodynamics: Phase transitions in ``Small''
  systems}, vol.~66 of \emph{Lecture Notes in Physics}, World Scientific,
  Singapore, 2001.

\bibitem{gross186}
Gross, D., \emph{PCCP}, \textbf{4},
  863--872,http://arXiv.org/abs/cond--mat/0201235 ((2002)).

\bibitem{uffink01}
Uffink, J., \emph{cond-mat/0005327}.

\bibitem{einstein05d}
Einstein, A., \emph{Annalen der Physik}, \textbf{17}, 132 (1905).

\bibitem{zermelo97}
Zermelo, E., \emph{Wied.Ann.}, \textbf{60}, 392--398 (1897).

\bibitem{boltzmann1896}
Boltzmann, L., ``Entgegnung auf die w\"armetheoretischen
Betrachtungen
  des Hrn.E. Zermelo'', in \emph{Kinetic Theory}, edited by S.~Brush, Pergamon
  Press, Oxford, 1965-1972, 2, p. 218.

\bibitem{gallavotti99}
Gallavotti, G., \emph{Statistical Mechanics}, Texts and Monographs in Physics,
  Springer, Berlin, 1999.

\bibitem{gaspard97}
Gaspard, P., \emph{J.Stat.Phys}, \textbf{88}, 1215--1240 (1997).

\bibitem{lebowitz99}
Lebowitz, J., \emph{Physica A}, \textbf{263}, 516--527 (1999).

\bibitem{lebowitz99a}
Lebowitz, J., \emph{Rev.Mod.Phys.}, \textbf{71}, S346--S357 (1999).

\bibitem{gibbs02}
Gibbs, J., \emph{Elementary Principles in Statistical Physics}, vol.~II of
  \emph{The Collected Works of J.Willard Gibbs}, Yale University Press
  1902,also Longmans, Green and Co, NY, 1928.

\bibitem{gibbs36}
Gibbs, J., \emph{Collected works and commentary,vol I and II}, Yale University
  Press, 1936.

\bibitem{gross183}
Gross, D., ``Ensemble probabilistic equilibrium and
non-equilibrium
  thermodynamics without the thermodynamic limit'', in \emph{Foundations of
  Probability and Physics}, edited by A.~Khrennikov, ACM, World Scientific,
  Boston, 2001, PQ-QP: Quantum Probability, White Noise Analysis XIII, pp.
  131--146.

\bibitem{falconer90}
Falconer, K., \emph{Fractal Geometry - Mathematical Foundations and
  Applications}, John Wiley \& Sons, Chichester, New York, Brisbane,
  Toronto,Singapore, 1990.

\bibitem{gross187}
Votyakov, E., Hidmi, H., Martino, A.~D., and Gross, D., \emph{Phys.Rev.Lett.},
  \textbf{89}, 031101--1--4; http://arXiv.org/abs/cond--mat/0202140 (2002)).

\end{thebibliography}


\end{document}